\documentclass{emulateapj}

\slugcomment{Draft version \today}

\shorttitle{Dust production in 47 Tuc}
\shortauthors{McDonald et al.}

\begin{document}

\title{Dust production and mass loss in the Galactic globular cluster 47 Tucanae}

\author{I.~McDonald\altaffilmark{1}, M.~L.~Boyer\altaffilmark{2}, J.~Th.~van Loon\altaffilmark{3}, A.~A.~Zijlstra\altaffilmark{1}}
\altaffiltext{1}{Jodrell Bank Centre for Astrophysics, Alan Turing Building, Manchester, M13 9PL, UK; iain.mcdonald-2@manchester.ac.uk}
\altaffiltext{2}{STScI, 3700 San Martin Drive, Baltimore, MD 21218, USA}
\altaffiltext{3}{Lennard-Jones Laboratories, Keele University, ST5 5BG, UK}

\begin{abstract}
Dust production among post-main-sequence stars is investigated in the Galactic globular cluster 47 Tucanae (NGC 104) based on infrared photometry and spectroscopy. We identify metallic iron grains as the probable dominant opacity source in these winds. Typical evolutionary timescales of AGB stars suggest the mass-loss rates we report are too high. We suggest that this is because the iron grains are small or elongated and/or that iron condenses more efficiently than at solar metallicity. Comparison to other works suggests metallic iron is observed to be more prevalent towards lower metallicities. The reasons for this are explored, but remain unclear. Meanwhile, the luminosity at which dusty mass loss begins is largely invariant with metallicity, but its presence correlates strongly with long-period variability. This suggests that the winds of low-mass stars have a significant driver that is not radiation pressure, but may be acoustic driving by pulsations.
\end{abstract}

\keywords{stars: mass-loss --- circumstellar matter --- infrared: stars --- stars: winds, outflows --- globular clusters: individual (NGC 104) --- stars: AGB and post-AGB}


\section{Introduction}
\label{IntroSect}

Stellar mass loss is of critical importance to the later stages of stellar evolution. Stars evolving up the red giant branch (RGB) lose mass via stellar winds. The remaining mass of a star leaving the RGB tip is the primary determinant of its position on the horizontal branch (HB; e.g.\ \citealt{Rood73,Catelan00}). Subsequent mass loss on the asymptotic giant branch (AGB) ejects the star's entire hydrogen envelope, creating a post-AGB star and (perhaps) a planetary nebula (PN). The timescale and end-point of AGB evolution may therefore not be determined by the nuclear burning rate, but by mass loss \citep{vLGdK+99,BSvL+09}. Winds from these stars are the most significant producers of interstellar dust and therefore provided the heavy elements present in Population I stars, including the Sun and Solar System \citep{Gehrz89,Sedlmayr94,Zinner03}.

Debate exists around many finer points of the mass loss process, particularly its variation with fundamental parameters such as luminosity, temperature and metallicity. Firstly, the amount of mass loss that occurs on the RGB remains largely unmeasured. RGB mass loss appears to determine the `second parameter' (after metallicity) that defines the morphology of HBs in globular clusters (GCs) \citep{FPB97,LC99,Catelan00}. Secondly, it is not proven where dust formation stars, and whether RGB stars can form significant amounts of dust in their ejecta. If many RGB stars do produce dust, it would add a large number of dust factories ejecting dust into a galaxy's interstellar medium (ISM) (cf.\ the missing dust source in many galaxies; e.g.\ \citealt{MBZ+09}). \citet{ORFF+07} suggest that dust production occurs episodically over a large portion of the RGB, while \citet{MvLD+09} have presented evidence that dust production is confined to the RGB tip, though gaseous mass loss is still present all along both the RGB and AGB \citep{MvL07,MAD09,DSS09}. Dust can also help drive a wind through radiation pressure on grains (e.g.\ \citealt{Lewis89}), though it is not certain whether winds which lack high-opacity amorphous carbon can be accelerated in this manner \citep{Willson00,Woitke06b,Hoefner07,LZ08,MSZ+10}. Dustless stars typically lose mass via magneto-acoustic processes in their chromospheres \citep{HM80,DHA84,MvL07}. Finally, the composition of circumstellar dust and the chemical processes by which it forms are only partly defined. Of particular interest is the r\^{o}le that metallic iron grains play in dust formation, and how metallicity affects dust production and composition \citep{MSZ+10}.

In this study, we aim to address these points. To do this, we explore dust production in the Galactic globular cluster 47 Tucanae (NGC 104). Globular clusters are unique laboratories with which we can study late-stage evolution of low-mass stars: globular clusters contain stars that are among the oldest in the Universe, and most clusters are comprised of a single (or at least dominating) stellar population with little internal variance in age or chemical composition. Comparisons between clusters therefore probe differences in mass loss created by metallicity, while comparisons within clusters probe differences created by stellar temperature and luminosity. Throughout this paper, we will refer to similar studies: \citet{MvLD+09}, which covers the cluster $\omega$ Centauri (NGC 5139; hereafter Paper I); \citet{BMvL+09}, which covers the cluster NGC 362 (hereafter Paper II); and a work which accompanies this paper (submitted ApJS; hereafter Paper III).

The cluster 47 Tuc itself is among the Galaxy's most massive and, co-incidentally, one of the closest to us. Its distance has been estimated photometrically as 4190--4700 pc \citep{Harris96,PSvWK02,SHO+07}\footnote{The updated Harris cluster catalogue can be found at: http://www.physics.mcmaster.ca/\protect\~{}harris/mwgc.dat} and kinematically as 4000 pc \citep{MAM06}, with 4500 pc being the median distance determination in the literature. Its mass is 6--9 $\times 10^5$ M$_\odot$ \citep{SR75,MM85,MSS91}. Its metallicity is determined to be roughly [Fe/H] = --0.7, though various studies have placed it anywhere from [Fe/H] = --0.89 to --0.66 \citep{Harris96,CG97,MB08,WCMvL10}. The cluster suffers from little interstellar reddening ($E(B-V) = 0.04$ mag --- \citealt{Harris96}) and is estimated to be over 11 Gyr in age \citep{CSD92,ZRO+01,GSA02,PSvWK02,GBC+03,KTR+07,SHO+07}. The cluster hosts an ionised intra-cluster medium (ICM; \citealt{FKL+01}). This ionised medium is undetected in any other cluster (M15 has a neutral medium with a dusty component; \citealt{ESvL+03,vLSEM06,BWvL+06}). Integrating the observed intracluster electron density over the cluster core yields $\sim$0.1 M$_\odot$ of H {\sc ii}, representing $\gtrsim$5\% of the intracluster hydrogen \citep{SWFW90,FKL+01}. The total stellar mass loss from 47 Tuc's stars therefore gives us insight into the fate of gas and dust expelled by stars in this cluster.

Paper III more-precisely examines which stars have mid-infrared excess, indicative of dust, finding again that we cannot corroborate the presence of dust around the fainter giants. In this work, we use the data and results from Paper III to derive mass-loss rates for individual stars and (by summation) the cluster as a whole, and investigate the composition of the circumstellar dust produced by analysing archival mid-infrared (mid-IR) spectra.


\section{Mass loss and dust production}
\label{SectMdot}

\subsection{The stellar mass-loss rate in 47 Tuc}

\subsubsection{The input data}

The data for this study come from the accompanying Paper III, which identifies those stars in 47 Tuc which have infrared excess, indicative of circumstellar dust. This list of 22 objects is an exhaustive list of those stars in the cluster which we can be relatively certain are producing dust, given currently available data. These objects are listed in Table \ref{MdotTable}, which also includes the parameters we determine in later sections.

\begin{center}
\begin{table*}
\caption{Calculated wind properties and literature pulsation properties for potentially dusty stars in 47 Tuc.}
\label{MdotTable}
\begin{tabular}{@{}l@{}c@{}c@{}c@{\ \ }c@{\ \ }c@{\ \ \ \ }c@{\ \ }c@{\ \ }c@{\ }c@{\ \ }c@{}c@{\ \ }c@{\ \ \ }c@{\ }c@{\ }c@{\ }cl@{}}
    \hline \hline
Star	& $\delta V^{(1)}$ & $\delta v^{(1)}$ & $P^{(1)}$ & $T_{\ast}$ & $L$ & \multicolumn{3}{c}{Obs./modelled flux} & $f_{\rm r}$ & $\dot{M}^{(2)}$ & $v_\infty^{(2)}$ & $T_{\rm inner}^{(3)}$ & \multicolumn{4}{c}{Contribution (\%)$^4$} & Notes \\
\ 	& (mag) & (km/s) & (days) & (K) & (L$_\odot)$ & $K_{s}$	& [3.6]	& [8]	& (\%) & \llap{(}10$^{-7}$\,M$_\odot$\,yr$^{-1}$) & (km\,s$^{-1}$\rlap{)} & (K) & Sil. & Fe & Al$_2$O$_3$ & FeO & \  \\
    \hline
V1     & 4.03  & 20    & 221   & 3623 & 4824 &0.936 & 1.173 & 2.213 & 3.6 & 21.0	& 4.0	& 900	& 12	& 88	& 0	& 0	& \\
V8     & 1.6   & 16    & 155   & 3578 & 3583 &0.904 & 1.210 & 2.135 & 5.9 & 14.7	& 4.0	& 900	& 7	& 85	& 7	& 1	& \\
V2     & 2.78  & 23    & 203   & 3738 & 3031 &1.103 &\nodata&\nodata& 4.3 & 12.0	& 3.8	& 900	& 5	& 95	& 0	& 0	& $^{5}$\\
V3     & 4.15  & 22    & 192   & 3153 & 2975 &0.954 & 1.198 &\nodata& 3.2 & 9.4	& 3.2	& 1000	& 0	& 100	& 0	& 0	& $^{6}$\\
V4     & 1.5   & 18    & 165   & 3521 & 2603 &0.932 & 1.167 & 1.902 & 5.6 & 11.7	& 3.6	& 900	& 2	& 95	& 2	& 1 	& $\sim$RGB-tip\\
V26    & 0.3   &\nodata&\nodata& 3500 & 2541 &1.073 & 0.807 & 0.995 & 0.9 & 5.9	& 3.8	& 1000	& 	& 100	& 	&  	& $^{7}$\\
LW10   & 1.2   &\nodata& 221:  & 3543 & 2324 &1.073 & 0.876 & 1.017 & 0.6 & 4.2	& 3.2	& 1000	& 0	& 100	& 0	& 0	& \\
V21    & 1.0   & 7     & 76:   & 3575 & 2301 &0.977 & 1.083 & 1.355 & 1.1 & 4.9	& 3.6	& 1000	& 0	& 93	& 5	& 2	& \\
LW9    & 1.1   &\nodata& 74    & 3374 & 2204 &0.973 & 1.003 & 1.228 & 1.1 & 6.0	& 3.3	& 950	& 	& 100	& 	&  	& \\
V27    & 0.3   &\nodata& 69    & 3374 & 2140 &1.023 & 0.929 & 1.218 & 1.4 & 3.3	& 3.6	& 1050	& 	& 100	& 	&  	& \\
Lee1424&\nodata&\nodata& 65    & 3565 & 2122 &1.006 & 1.182 & 1.390 & $>$4& 4.4	& 4.0	& 1050	& 	& 100	& 	&  	& \\
A19    & 0.7   &\nodata& 60    & 3526 & 2096 &1.034 & 0.992 & 1.240 & 0.8 & 3.6	& 4.2	& 1100	& 	& 100	& 	&  	& \\
LW12&0.6\rlap{$^8$}&\nodata&116& 3713 & 2079 &1.082 & 0.967 & 1.093 & $<$1& low\rlap{$^9$}\\
x03    &\nodata&\nodata&\nodata& 3816 & 1640&\nodata& 0.958 & 1.156 & 3.2 & 5.1	& 1.8	& 750	& 	& 100	& 	& 	& $^{11}$\\
LW19   & 0.1   &\nodata& 40    & 3738 & 1638 &1.078 & 0.991 & 1.080 & $<$1& low\\
V20    & 0.8   &\nodata& 232:  & 3602 & 1575 &1.053 & 0.928 & 1.039 & $<$1& low\rlap{?}&&&&&&&$^{10,11,12}$\\ 
V22    & 0.5   &\nodata& 62    & 3684 & 1528&\nodata& 2.504 & 2.834 & $<$1& low	&	&	&	&	&	&	& $^{7}$\\
V6     & 0.6   & 7     & 48    & 3763 & 1374 &1.001 & 1.000 & 1.167 & $<$1& low&&&&&&&\\	
V5     & 0.7   & 8     & 50    & 3741 & 1363 &1.003 & 1.100 &\nodata& $<$1& low&&&&&&&$^{7}$\\	
V18    & 0.3   & 5     & 83:   & 3692 & 1297 &0.989 & 1.004 &\nodata& 2.4  & 5.3	& 2.4	& 800	& 33	& 67	& 0	& 0	& $^{11}$\\
V23    & 0.5   &\nodata& 52    & 3775 & 1259 &1.023 & 0.985 &\nodata& $<$1& low&&&&&&&$^{7}$\\
V13    &\nodata& 12    & 40    & 3657 & 1029 &0.990 & 1.106 &\nodata& 1.2 & 4.1	& 2.2	& 800	& 0	& 85	& 15	& 5	& $^{11}$\\
   \hline
\multicolumn{18}{p{0.95\textwidth}}{$^{1}$Optical and radial velocity variability (optical in $V$-band, where known) with periods (colons denote approximate values, preference given to the longest specified period), from \citet{Clement97}; \citet{Pojmanski02}; \citet{LW05}; and \citet{LWH+05}. $^2$Total (gas\,+\,dust) mass-loss rate and terminal velocity assuming a purely radiatively-driven wind, assuming $\psi = 1/1000$. $^3$Temperature of the inner edge of the circumstellar dust envelope. $^4$Implied fraction of grains in `astronomical' silicates, metallic iron, aluminium oxide and iron oxide. $^5$No \emph{Spitzer} IRAC data, fit based on literature photometry and MIPS 24-$\mu$m data. $^6$Unusually cool object. $^7$Photometry uncertain due to blending, fits made to \emph{Spitzer} IRAC data only. $^8$May switch between two states of variability, see \citet{LW05}. $^9$Values titled low are undetermined, but liable to be $\lesssim 2 \times 10^{-7}$ M$_\odot$ yr$^{-1}$. $^{10}$8- \& 24-$\mu$m emission only, suggesting the presence of silicate grains. $^{11}$Wind may be impossible to sustain (see text). $^{12}$Very poor-quality $JHK_{s}$ photometry, IR excess difficult to determine.}\\
    \hline
\end{tabular}
\end{table*}
\end{center}

\subsubsection{A measure of mass-loss rate}

Accurate --- or even approximate --- mass-loss rates are notoriously difficult to measure. One must assume that the wind has certain properties. Perhaps the most crucial of these is the relationship between wind velocity and distance from the star. We here assume a constant, spherically-symmetric, purely radiatively-driven wind, starting from a near-stationary, dust-forming layer. As discussed previously (Paper I; \citealt{MSZ+10}), this may not be the most appropriate description, but suffices due to lack of contrary evidence: we return to this point in the discussion.

Before calculating an estimated mass-loss rate, we can create a measure of the amount of circumstellar dust surrounding a star, irrespective of most modelling uncertainties. To do this, we define a quantity $f_{\rm r}$, the fraction of the star's luminosity that is reprocessed by circumstellar dust. For low optical depths, this should be directly proportional to the surface area of dust grains, therefore approximately proportional to the mass of circumstellar dust. The value of $f_{\rm r}$ can be determined by taking the difference between the observed SED and model photospheric SED in the infrared (yielding the infrared excess) and dividing this by the total photospheric luminosity (the value given in Paper III). We list this quantity in Table \ref{MdotTable}. Errors in the value of $f_{\rm r}$ are introduced by difficulties in determining the integrated luminosity from photometry alone and from the uncertainty in the photospheric spectrum (both temperature and luminosity). The former can be approximated as $\Delta f_{\rm r} = \sqrt{1+f_{\rm r}}-1$, or roughly 50\% of $f_{\rm r}$, but is dominated by photometric uncertainty at short (2.2--10-$\mu$m) wavelengths. The latter gives a constant uncertainty of around $\Delta f_{\rm r} \approx 0.4$\%. We reduce this latter error by fixing the photospheric luminosity to the $K$-band flux for all stars except V26 and x03, which are fixed to the 3.6-$\mu$m flux. The value $f_{\rm r}$ could not reliably be measured for several stars with low mass-loss rates.

The stars appear to fall into three main categories: those with large ($\sim$5\%) reprocessed flux (V1--4, V8), those with low ($\sim$2\%) reprocessed flux concentrated toward long wavelengths (V13 and V18), and those stars with low ($\lesssim$2\%) reprocessed flux concentrated toward shorter wavelengths (the remainder).

\subsubsection{Defining the mass-loss rate}

Given the limitations of our understanding and of the model presented above, we stress that the mass-loss rates and velocities we derive are based only on one model for wind acceleration, but should be comparative both internally and with other works which use these same assumptions.

Under these assumptions, we calculate mass-loss rates by modelling the SED using the radiative transfer package, {\sc dusty} \citep{NIE99}. As inputs, we use our interpolated {\sc marcs} model of the stellar spectrum, and a standard MRN grain size distribution \citep{MRN77}. Within our {\sc dusty} model, we vary the dust composition, temperature in the dust-forming region (the inner edge of the dust envelope) and optical depth of the wind. We change these parameters iteratively until a best fit is achieved.

We relate the {\sc dusty} mass-loss rate ($\dot{M}_{\rm DUSTY}$) to the actual mass-loss rate ($\dot{M}$) as previously (Paper I; Paper II), using the following formula \citep{vanLoon00}:
\begin{eqnarray}
    \dot{M} &=& \dot{M}_{\rm DUSTY} \left(\frac{L}{10^4\ {\rm L}_\odot}\right)^{3/4}
		      \left(\frac{200}{\psi} {\rm \ }
			\frac{\rho}{3 \ {\rm g\,cm}^{-3}}\right)^{1/2}\\ \label{MdotEq1}
	    &=& 10^{-3} \dot{M}_{\rm DUSTY}L/v_{\infty}\quad [{\rm M}_\odot\ {\rm yr}^{-1}], \label{MdotEq2}
\end{eqnarray}
for luminosity $L$, dust-to-gas ratio $\psi$, grain density $\rho$, and terminal wind velocity $v_{\infty}$. We assume that $\rho = 3$ g cm$^{-3}$ for silicates, 4 g cm$^{-3}$ for aluminium oxides, 5.7 g cm$^{-3}$ for FeO, and 7 g cm$^{-3}$ for metallic iron. For these three species, we use the optical constants from \citet{DL84}; \citet{BDH+97}; and \citet{OBA+88}, respectively. The errors in the above assumptions are substantial, and are discussed in Section \ref{MdotErrorSect}.

The dust-to-gas ratio, $\psi$, can be estimated as:
\begin{equation}
	\psi = \frac{1}{200} \times 10^{\rm [Fe/H]} ,
  \label{PsiEqn}
\end{equation}
which is approximately 1/1000 for 47 Tuc. This assumes that the fraction of condensed metals is identical for all metallicities. We discuss the accuracy of this assumption in Section \ref{MdotErrorSect}.

\subsubsection{The stellar mass-loss rate}
\label{MdotSect}

\begin{figure*}
\includegraphics[width=0.7\textwidth,angle=-90]{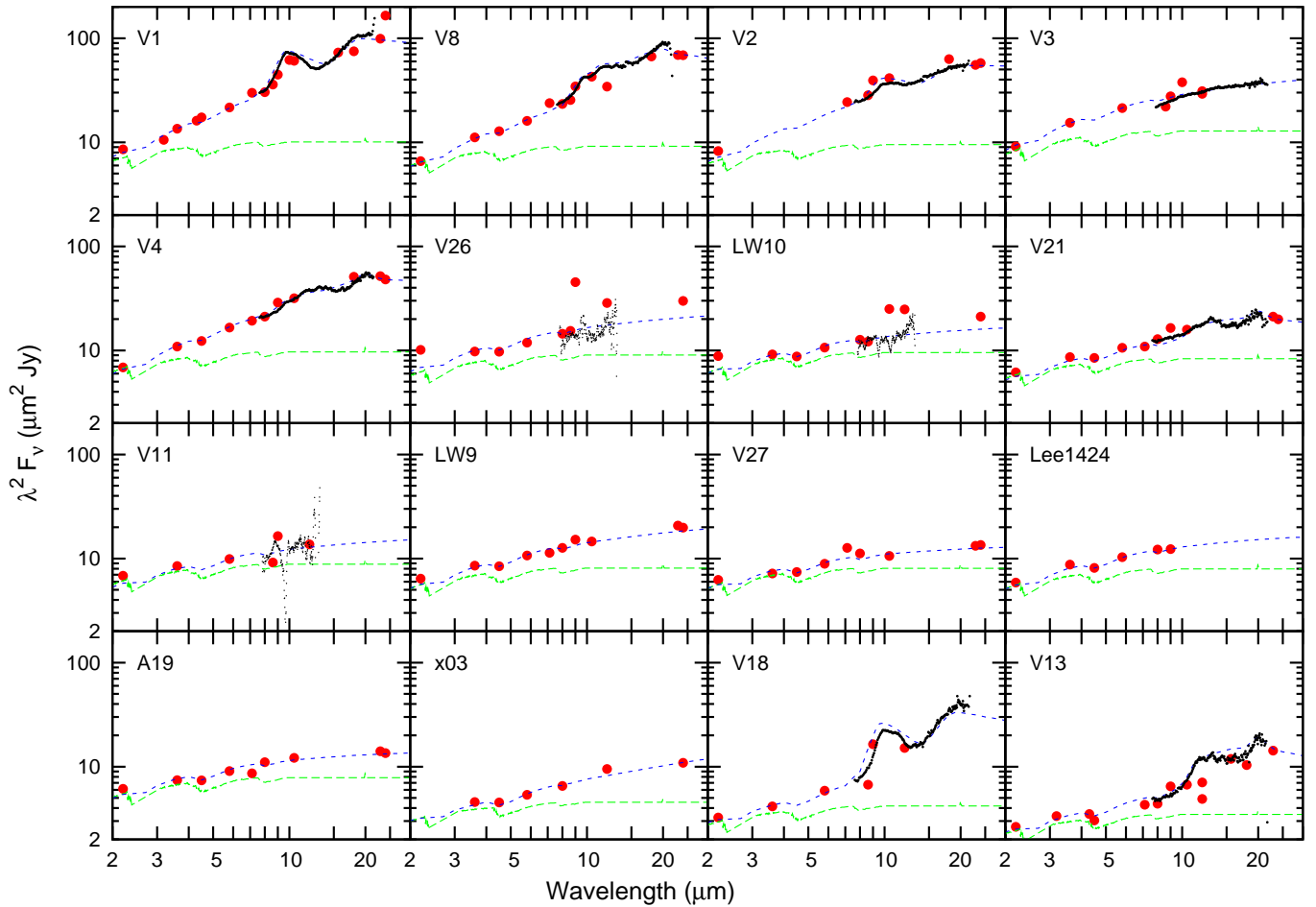}
\caption{IR portions of the SEDs of dusty stars, plotted in Rayleigh--Jeans units. Large (red) and small (black) points show literature photometric and smoothed spectroscopic data, respectively (see \S\ref{MdotSect} for reference list). Spectroscopic data have been scaled in flux to match the photometric data. The long-dashed (green) and short-dashed (blue) lines show the modelled photospheric contribution and total flux, respectively.}
\label{SEDsFig}
\end{figure*}

Table \ref{MdotTable} shows our calculated mass-loss rates and inferred wind parameters and dust compositions alongside literature pulsation properties. The striking fact about the dust composition is the fraction of iron grains present. The presence of metallic iron in these stars is debatable, as its featureless infrared emission can be mimicked by many other phenomena, including large silicate grains and emission from thick circumstellar molecular layers \citep{MSZ+10}. We consider iron to be the most likely source of emission, though the fractions of metallic iron listed in Table \ref{MdotTable} may not be particularly accurate, and we discuss this further in \S\ref{DustCompSect}.

Many of the mass-loss rates are too low to be sensibly measured: we can merely say that some circumstellar dust probably exists. Those rates that can be measured have their SEDs shown in Figure \ref{SEDsFig}. The photometry is taken from the sources cited in Paper III\footnote{Optical data: \protect\citet{Stetson00,ZHT+02,SHO+07}; near-IR data: \protect\citet{SCS+06,SHO+07}; mid-IR data: \protect\citet{OSA+97,OFFPR02,vLMO+06,ITM+07,BBW+09,GMB+09,BvLM+10,KAC+10}}. Mid-IR spectra are included from \citet{vLMO+06} (hereafter vLMO), and \citet{LPH+06} (hereafter LPH). These objects are summarised in the following luminosity-ordered list:
\begin{list}{\labelitemi}{\leftmargin=1em \itemsep=0pt}
\item V1: This star is clearly identified as the cluster's dominant dust producer by vLMO, LPH and \citet{ITM+07} (hereafter ITM). V1 has clear excess at all wavelengths $> 3 \mu$m. This excess, which we attribute to metallic iron dust, dominates even over its strong silicate feature.
\item V8: No silicate feature was seen in this star by vLMO, but a clear silicate and aluminium oxide feature was seen by LPH. It has previously been suggested (LPH; Paper I) that V8 may have variable mass loss. However, the \emph{AKARI} photometry at 3, 4, 7, 11, 18 and 23 $\mu$m shows no variation in the (dominant) iron contribution. This means that if the mass loss is variable, strong variation only occurs in the silicate production. V8 has a complex IR spectrum showing peaks from silicates, aluminium oxide, and from crystalline minerals at 9.7, 11.3, 13.1, 20 and 32 $\mu$m. We discuss the carrier(s) of these bands in \S\ref{DustCompSect}. Note that the amplitude of the 10-$\mu$m silicate peak is substantially over-estimated in our model compared to the 20-$\mu$m silicate peak. This implies a lower temperature for the silicate grains than the metallic iron grains, which we also discuss in \S\ref{DustCondSect}.
\item V2: A silicate dust feature was observed by LPH, though it was too weak to have been seen by vLMO. Iron dominates the IR flux, however. The 10- and 20-$\mu$m silicate peaks are not well modelled, as in V8 (see also \S\ref{DustCondSect}).
\item V3: No silicate emission was seen by either LPH or vLMO, but strong, featureless (metallic iron) emission is present, which is confirmed by \emph{AKARI} photometry.
\item V4: The spectrum in LPH shows complex dust mineralogy, akin to that in V8. This star is roughly at the RGB-tip, as estimated from the luminosity function.
\item V26: Blended with a star half its brightness, 2.2$^{\prime\prime}$ to the east, the near-IR, \emph{AKARI} and 24-$\mu$m photometry are affected. The amount of IR excess is difficult to determine here, but is likely to be quite small. No discernible features are present in the spectrum from vLMO. 
\item LW10: Also within the dense cluster core, it appears that the \emph{AKARI} photometry (at 9 \& 11 $\mu$m) for this object is affected. No features are detectable in the spectrum from vLMO.
\item V21: Observed by LPH, the spectrum shows the same features as V8. These features, and the silicate contribution, are considerably weaker in V21.
\item V11: No obvious features were seen in the spectrum of vLMO, this star was only covered by \emph{Spitzer} at 3.6 and 5.8 $\mu$m. \emph{AKARI} 9-$\mu$m photometry suggests it may have some excess there, but both the \emph{Spitzer} 8-$\mu$m photometry and the vLMO spectrum (which covers 9-$\mu$m) show no excess. The derived mass-loss rate can probably be treated as a rough upper limit.
\item LW9 \& V27: An excess consistent with circumstellar dust is present in both the \emph{Spitzer} and \emph{AKARI} photometry. The chemistry cannot be well-constrained by this photometry, but it appears that the flux contribution from components other than iron is comparatively small.
\item Lee1424: Lacks any data redwards of 9 $\mu$m, making constraining the mass-loss rate, dust temperature and dust chemistry very difficult. The values given should be treated as order-of-magnitude estimates only.
\item A19: As LW9 \& V27, however the \emph{AKARI} 7-$\mu$m data suggests little to no mid-IR excess in this star.
\item V20 \& V22: not shown in Figure \ref{SEDsFig}, they have no near-IR photometry attributed to them, meaning it is difficult to determine the amount of IR excess present. Despite this, the dust temperature is relatively-well constrained by the presence of \emph{Spitzer} 24-$\mu$m photometry.
\item x03: Located in the cluster core, this star suffers from the same problems as V20 \& V22, but clearly displays a stronger excess, making an approximate determination of the mass-loss rate and wind characteristics possible.
\item V18: Observed by both LPH and vLMO, this star has a clear silicate feature. This feature was not claimed in vLMO due to poor signal-to-noise, but the detection appears to match that from LPH well. It shows substantial mass loss compared to other stars of similar luminosity. The 10- and 20-$\mu$m silicate peaks are not well modelled, as in V8 (see also \S\ref{DustCondSect}).
\item V13: Highlighted in both LPH and ITM, the spectrum of this star shows an excess similar to that of V8. The contribution of iron dust is comparatively small, though would appear to be non-zero. This star is isolated from the cluster core, but has poor-quality photometry due to its comparative faintness.
\end{list}

\subsubsection{Error in the mass-loss rate}
\label{MdotErrorSect}

The accuracy of the wind parameters we find is hard to determine. The major sources of error are listed below:
\begin{enumerate}
\item A 30\% uncertainty inherent in {\sc dusty} due to an unspecified gravitational correction. This uncertainty is exceeded in stars with $v_\infty \lesssim 2.5$ km s$^{-1}$, where the effect of stellar gravity will change the variation of dust particle density with radius from the star (the stellar mass is unspecified in {\sc dusty}).
\item Uncertainty in the underlying photospheric spectrum. As the iron dust contribution is essentially a modified blackbody, one can fit a dust temperature to the 3.6- to 8-$\mu$m data and extrapolate this fit backwards to $K_{s}$-band, where the dust contribution is essentially zero. The photospheric contribution can then be set from the $UBVIJHK_{s}$ photometry, but the accuracy of this fit depends on the stellar photometry. (This would equally hold for other possible dust species, e.g.\ amorphous carbon.)
\item Uncertainty in the amount of IR excess. Apart from identifying how much flux is generated by the photosphere, missing photometry and photometric errors cause uncertainty in the combined flux of the (photosphere + dust) contributions.
\item Uncertainty in the dust composition. It is not clear for a number of these stars (V26, LW9, V27, Lee1424, A19, x03 and any star marked ``low'') what the dust chemistry is. Many of these stars have insufficient coverage in the 9--23 $\mu$m region to determine whether they contain any dust components other than metallic iron. LW10 and V11 are not covered by LPH: ground-based spectra by vLMO show no obvious spectral features, so we presume a pure metallic iron wind, but the spectra are of low signal-to-noise. The presence of silicates can significantly increase the implied mass-loss rate due to their low opacity. Any carbonaceous dust present would also be hidden by the modified blackbody of metallic iron. Carbonaceous dust typically has a higher opacity: if it is present, the implied mass-loss rate would be decreased.
\item Uncertainty in grain size distribution and grain shape (assumed here to be spherical, with radius $a$). The relationship between the grains' absorption cross-section to radiation ($\propto a^2$) and their mass ($\propto a^3$). A high grain porosity would also imply a decreased dust density and therefore a higher mass-loss rate. Particularly, differences in grain properties between the iron and silicate grains could explain why we model such a high proportion of iron grains in the wind (see \S\ref{DustCompSect}).
\item Uncertainty in dust temperature. The dust temperature can only be determined to within $\sim$100 K. As $L \propto R^2T^4$, a 15\% error in temperature leads to an error of up to $\sim$60\% in the flux emitted by the circumstellar dust, depending on the implied error in the radius of the dust shell. This will be most significant in stars that are losing the least mass, as their dust temperatures are more difficult to determine.
\item Uncertainty in the wind driving mechanism. We presume here that the winds are sustained by radiation pressure alone. The very low outflow velocities present in at least three cases (x03, V18 and V13) would suggest that there is insufficient radiation pressure for it to be the sole driver of mass loss. Additional energy input, for example from pulsation, would increase the wind velocity and (by virtue of Eq.\ (\ref{MdotEq2})) decrease the mass-loss rate. There is a related uncertainty in the dust-to-gas ratio, which is assumed to scale with metallicity. These uncertainties cannot be quantified statistically.
\end{enumerate}

\begin{center}
\begin{table*}
\caption{Error analysis of mass-loss rates for V1.}
\label{MdotErrorTable}
\begin{tabular}{lllll}
    \hline \hline
Parameter		& $\delta\dot{M}$ & $\delta v_{\infty}$ & Parameter variation used  	& Other parameters affected\\
   \hline
Grain size$^1$	& $\pm$10\%	& $\pm$30\%	& Low: $q = 2.5$, $r = 1$\,nm, $R = 100$\,nm		& $T_{\rm inner} = 750$ K, 90\% iron, 10\% silicates\\
\ 			& \ 		& \ 		& Assumed: $q = 2.5$, $r = 5$\,nm, $R = 250$\,nm	& $T_{\rm inner} = 850$ K, 88\% iron, 12\% silicates\\
\ 			& \ 		& \ 		& High: $q = 1.5$, $r = 5$\,nm, $R = 250$\,nm		& $T_{\rm inner} = 850$ K, 80\% iron, 20\% silicates\\
Grain density	& --50\%	& --50\% 	& Low: 1.63 g cm$^{-3}$				& \\
\ 			& \ 		& \ 		& Assumed (solid): 6.52 g cm$^{-3}$		& \\
Dust formation	& $\pm$10\%	& $\pm$20\% & Low: 750 K					& $\tau_{\rm V} = 0.38$, 92\% iron, 8\% silicates\\
\ \ temperature	& \ 		& \ 		& Assumed: 850 K					& $\tau_{\rm V} = 0.45$, 88\% iron, 12\% silicates\\
\ 			& \ 		& \ 		& High: 950 K					& $\tau_{\rm V} = 0.52$, 82\% iron, 18\% silicates\\
Dust:gas ratio	& $^{-55\%}_{+125\%}$	& $^{+125\%}_{-55\%}$ 	
							& Low: 1 / 5000					& \\
\ 			& \ 		& \ 		& Assumed: 1 / 1000				& \\
\ 			& \ 		& \ 		& High: 1 / 200					& \\
Velocity distribution & $\sim$+500\% 	& N/A 		
							& Assumed: 0 $\rightarrow$ v$_\infty = 4$ km s$^{-1}$ & $T_{\rm inner} = 850$ K\\
\ 			& \ 		& \ 		& High: 10 km s$^{-1}$ (constant)		& $T_{\rm inner} = 900$ K\\
Photosphere 	& $\pm$12\% & $\pm$2\% 	& Assumed $\pm$ 5\% in luminosity		& $\tau_{\rm V}$ decreased from 0.45 to 0.42\\
\ 			& $\pm$15\% & $\pm$6\% 	& Assumed $\pm$ 250 K in temperature	& \\
Calculation error$^2$ 	& $\pm$30\% & \  	& \ 		& \ \\
   \hline
Total			& $\sim^{+7}_{-4}\times$	& $\sim^{+N/A}_{-3.5}\times$	& \ 	& \ \\
   \hline
\multicolumn{5}{p{0.95\textwidth}}{$^{1}$The distribution of grain radii ($a$) is defined by a slope of $a^{-q}$ between sizes $r$ and $R$. $^2$Due to an unspecified correction for stellar gravity within \protect{\sc dusty} \protect\citep{NIE99}.}\\
\end{tabular}
\end{table*}
\end{center}

The easiest way to see how these uncertainties affect a star's determined mass-loss rate is to propagate them through the {\sc dusty} code. The results for V1 (Table \ref{MdotErrorTable} suggest that, in absolute terms, our mass-loss rates are little more than an educated guess. However, since a large number of other studies use the same assumptions, one can use these rates to trace differences between stars, providing these properties are largely invariant between stars. 

The parameters listed in Table \ref{MdotErrorTable} have different effects. The grain size, for example, has little effect on the mass-loss rate but significant effects on other wind properties. Conversely, the dust-to-gas ratio have significant effects on the mass-loss rate and velocity but little effect on the other wind parameters (temperature, optical depth and minerology). Clearly the largest effect on the mass-loss rate is whether the wind has an initial velocity driven by, for example, pulsation. Determining the radial velocity profile of the wind would reduce the error substantially. While this has not been done accurately for such low-metallicity, dust-producing giants, inferences from chromospheric line modelling (e.g.\ \citealt{JS91}; \citealt{MvL07}; \citealt{VMC+10}) appears to show that wind velocities do not decrease substantially from $\sim$10 km s$^{-1}$ in the stages shortly prior to dust formation. We point out, however, that the highest mass-loss rates (and by inference the highest wind velocities) may be ruled out on evolutionary grounds (\S\ref{DustCompSect}).

We emphasise that the errors derived above are not formal and are only estimates. These compare relatively well to the adopted error of $\sim$5$\times$ for silicate-dominated winds (e.g.\ \citealt{SSM+10}), where the grain properties are somewhat better determined due to the long history of their study and constraints from their spectral features. 

In relative terms, however, we can probably assume that the star-to-star variations are much smaller. In particular, the driving mechanism and hence the wind's initial velocity should not differ as markedly as the uncertainties we assume. Likewise, the dust-to-gas ratio will only depend on the condensation fraction in the wind, which should not vary as much as a factor of 5$\times$ between stars at the same metallicity. This reduces the overall error among stars in the same cluster to a factor of $\sim$2$\times$ and allows us to make direct comparisons between stars.


\section{Discussion}
\label{SectDisc}

\begin{figure}
\includegraphics[width=0.35\textwidth,angle=-90]{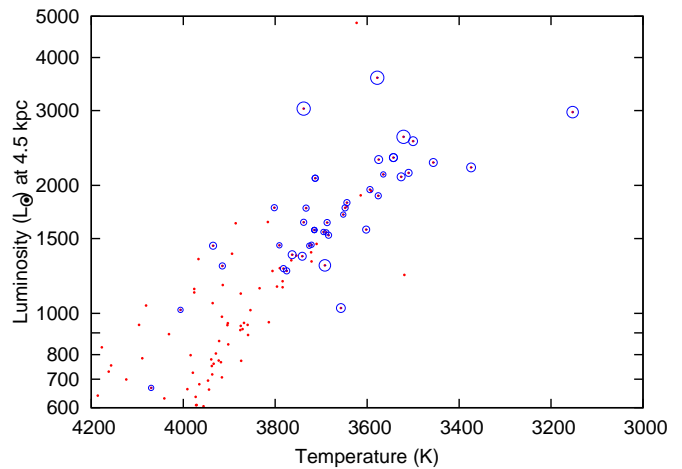}
\caption{Hertzsprung--Russell diagram, showing only the uppermost parts of the RGB and AGB. Known long-period variables are circled, with circle sizes corresponding to their mass-loss rates.}
\label{HRDVarsFig}
\end{figure}

\begin{figure}
\includegraphics[width=0.35\textwidth,angle=-90]{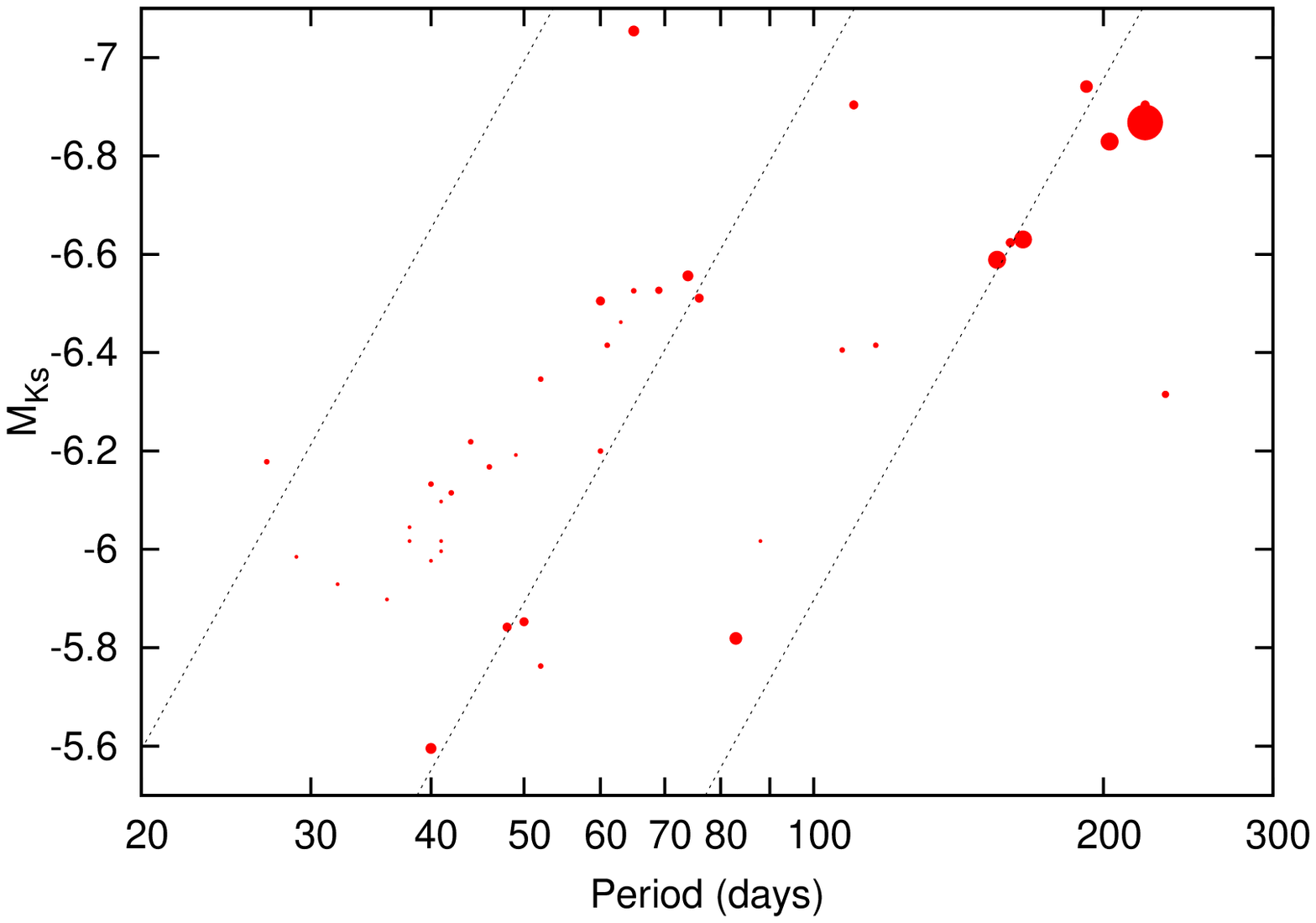}
\includegraphics[width=0.35\textwidth,angle=-90]{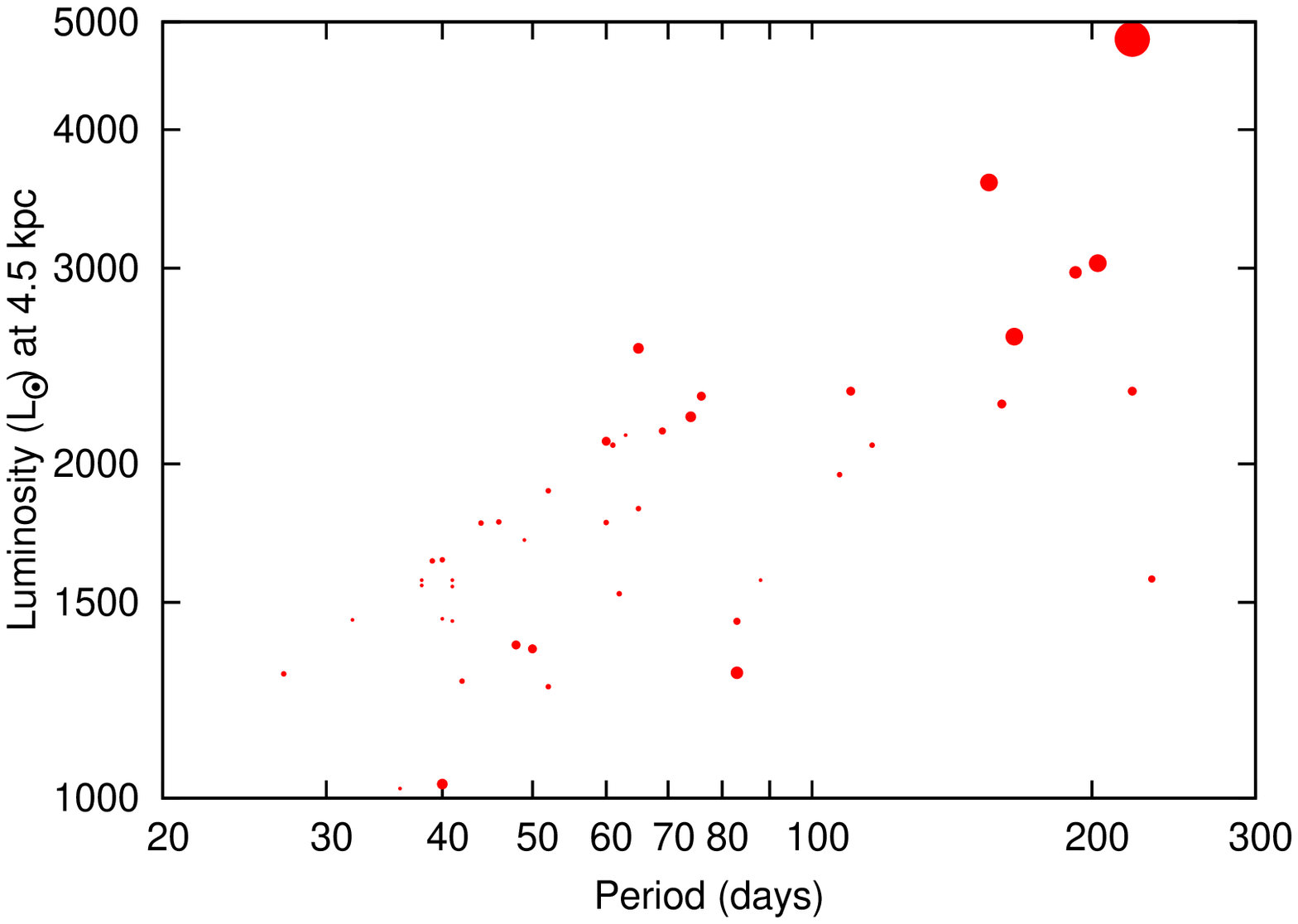}
\caption{Top panel: period--$M_{Ks}$ diagram for 47 Tuc's long-period variables, overplotted with relations derived for the Large Magellanic Cloud \citep{ITM+04}; bottom panel: period--luminosity diagram for the same stars. Dot size corresponds to mass-loss rate.}
\label{PLFig}
\end{figure}

\begin{figure}
\includegraphics[width=0.20\textwidth,angle=-90]{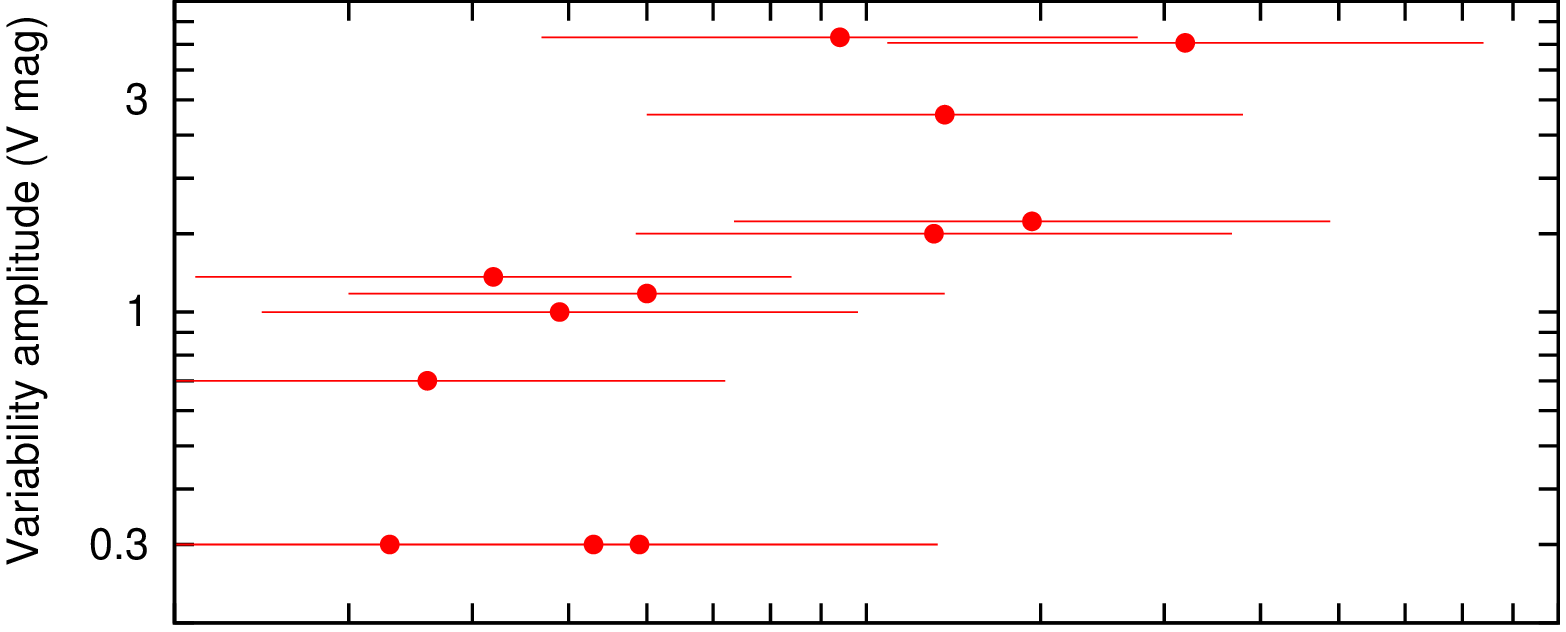}
\includegraphics[width=0.20\textwidth,angle=-90]{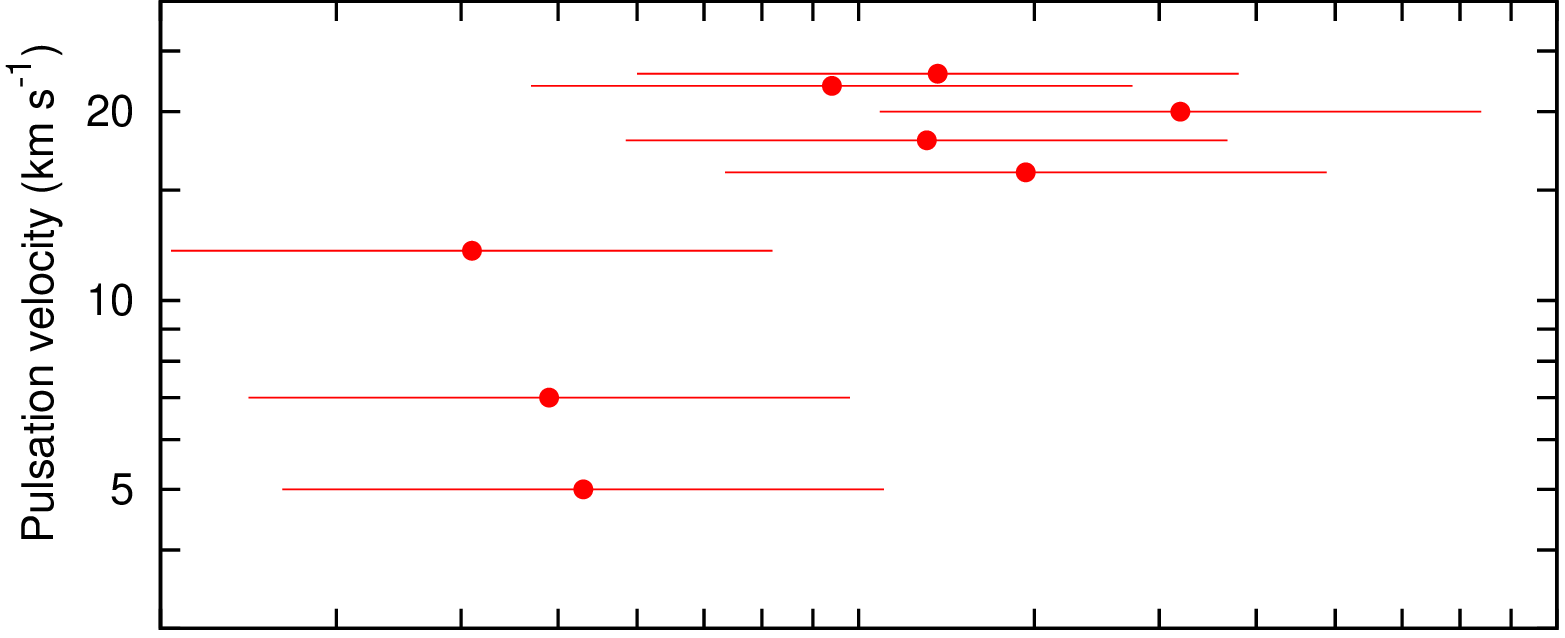}
\includegraphics[width=0.20\textwidth,angle=-90]{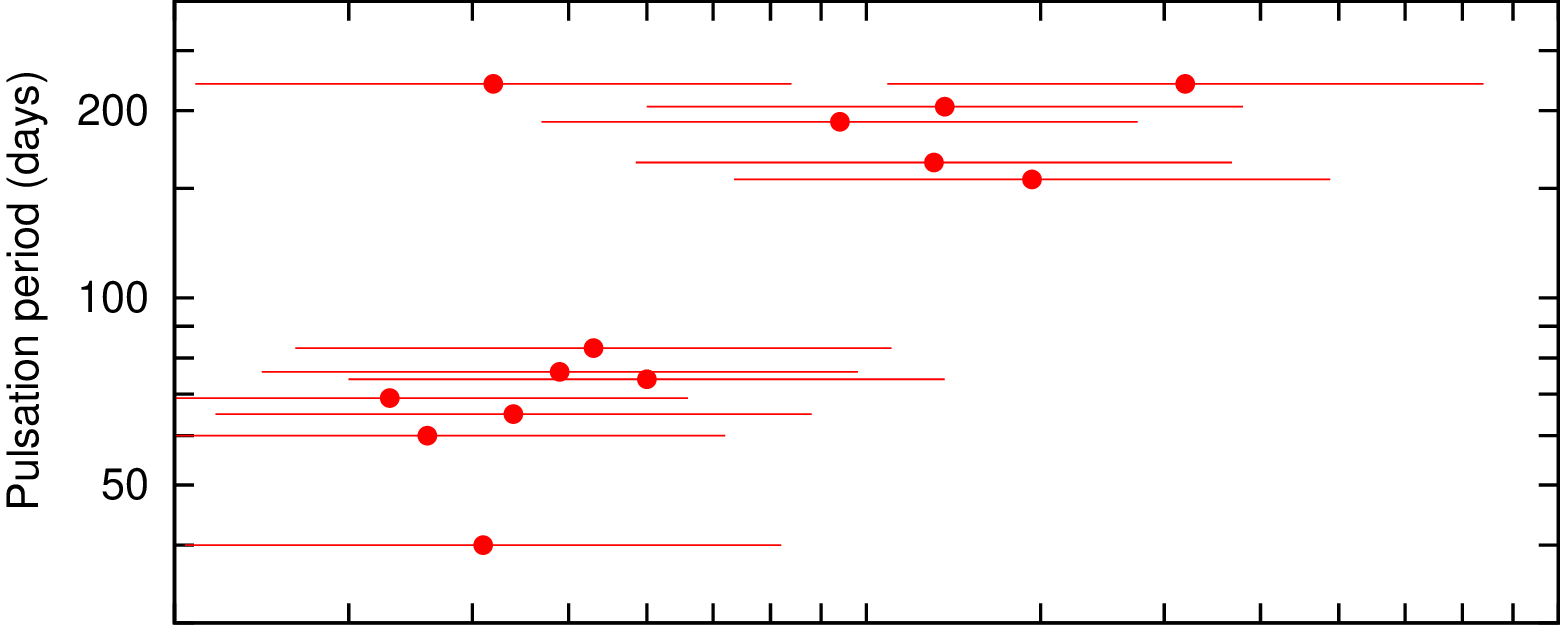}
\includegraphics[width=0.20\textwidth,angle=-90]{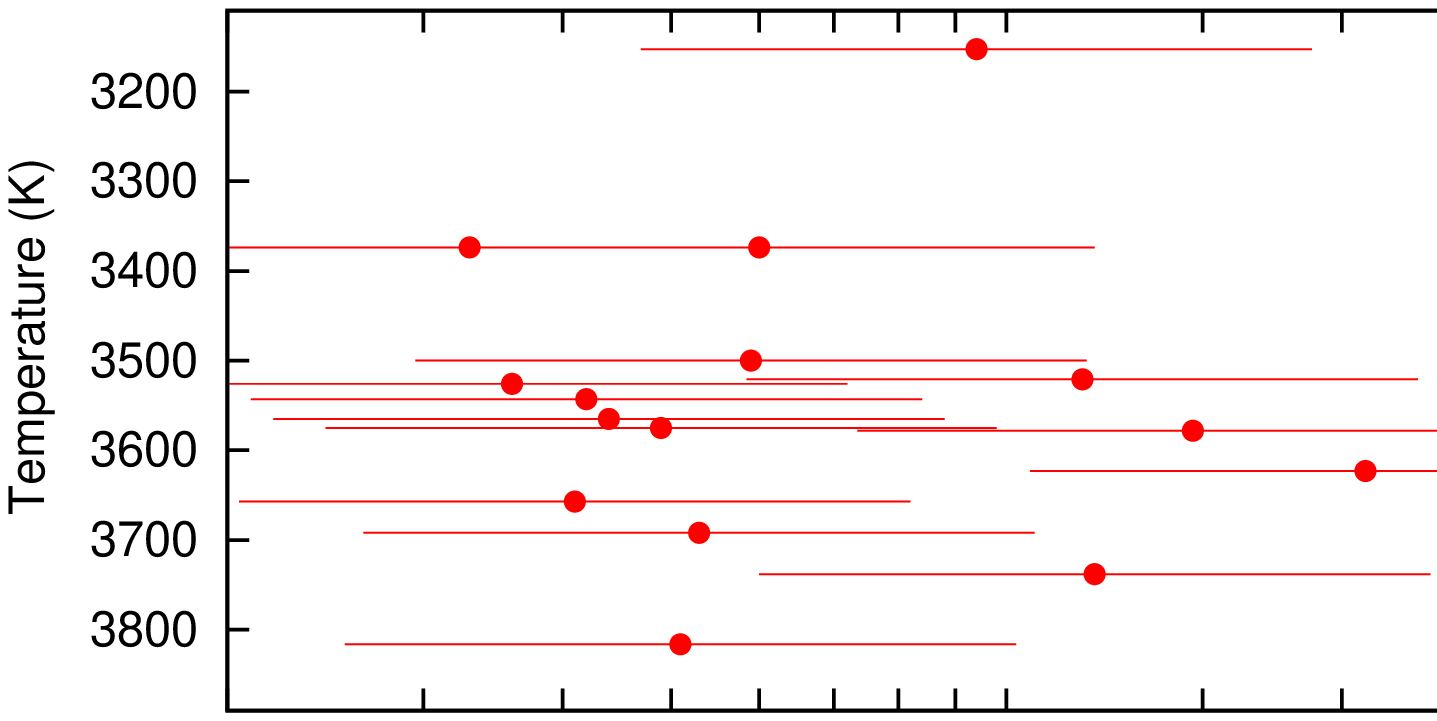}
\includegraphics[width=0.20\textwidth,angle=-90]{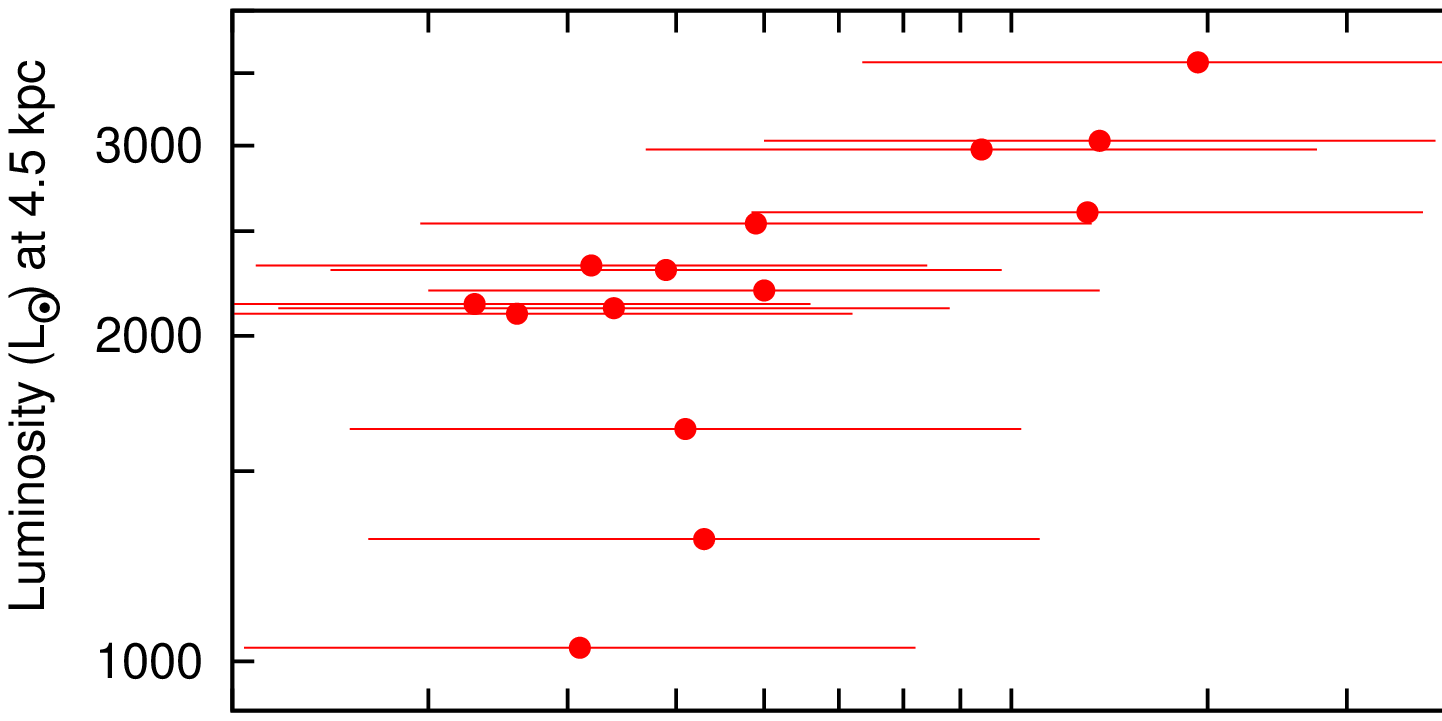}
\includegraphics[width=0.20\textwidth,angle=-90]{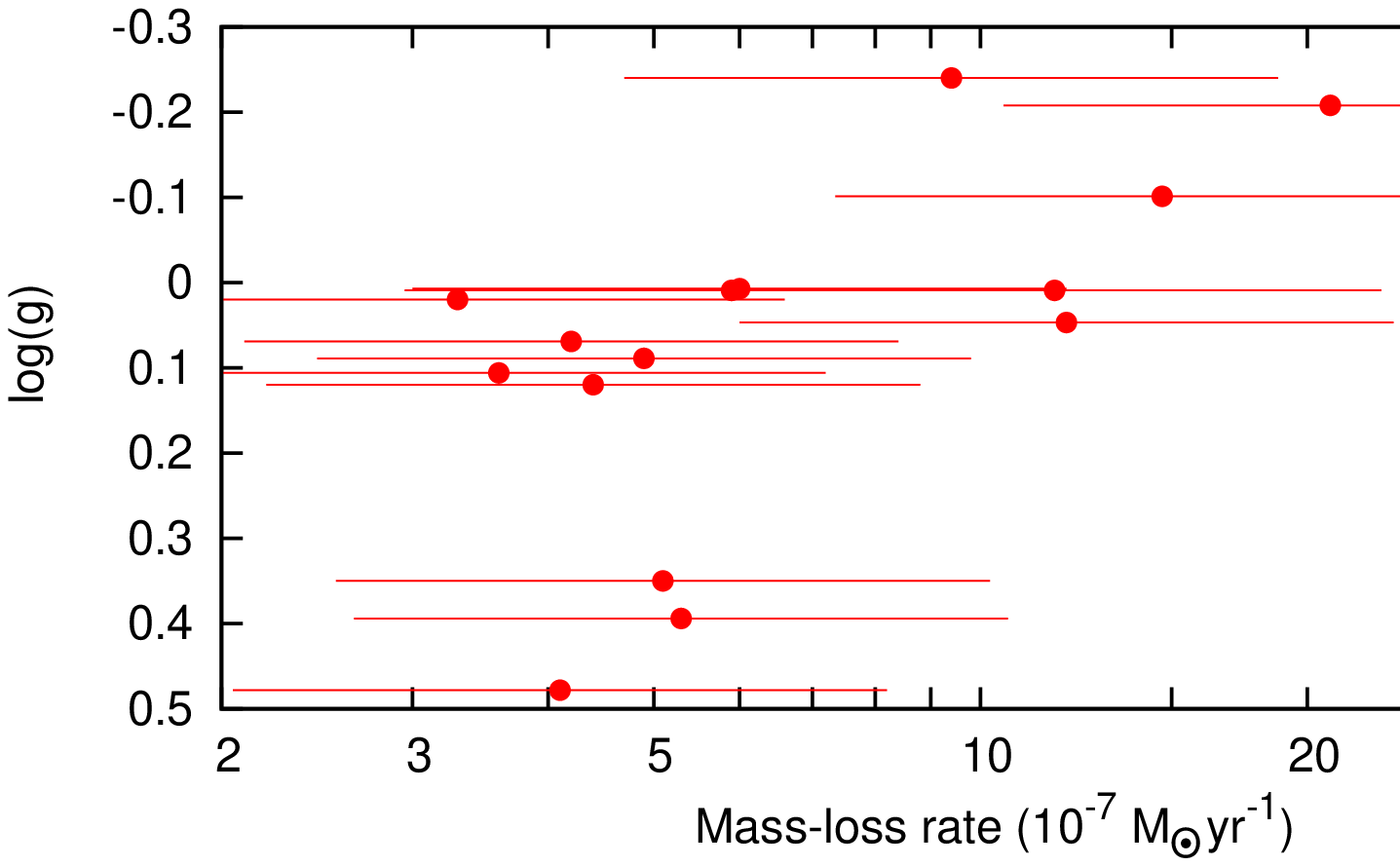}
\vspace{8mm}
\caption{Mass-loss rate \emph{versus} a variety of stellar parameters. Error bars of a factor of two are added, as discussed in \S\protect\ref{MdotErrorSect}.}
\label{MdotRelFig}
\end{figure}

The wind parameters listed in Table \ref{MdotTable} pose several problems for stellar wind theory, which we discuss in this Section. Firstly, the projected terminal velocities we derive are very low, implying radiation pressure from the star is not effective in driving the wind: we must consider whether other processes may contribute to, or dominate, the wind acceleration. Secondly, when one considers the evolutionary timescales of the stars in question, our derived mass-loss rates imply that the stars are losing much more material than is predicted from stellar theory. Finally, we can use these problems to constrain some of the wind parameters listed in Table \ref{MdotErrorTable}. 

Mass-loss rate, pulsation amplitude and pulsation period are all known to correlate with stellar evolution. This is demonstrated in Figures \ref{HRDVarsFig}, \ref{PLFig} and \ref{MdotRelFig}. These can be used to help solve the above problems.

\subsection{Efficiency of wind acceleration}
\label{WindAccSect}

A major problem when calculating mass-loss rates from metal-poor stars is that the terminal velocities predicted from radiative transfer codes are very low. This is a particular worry with {\sc dusty}, as no correction is input for stellar gravity. In this situation, it is impossible not to have mass lost from a star when dust is present, whereas the real situation may be that radiation pressure on dust is insufficient for it to overcome the stellar gravity field. Assuming the dust layers around these stars do have some outward velocity, a different method of wind acceleration may be required.

{\sc Dusty} provides a minimum mass for a radiative transfer model. If a star is less massive than this, the error due to the unspecified correction for stellar gravity will be greater than the 30\% inherent in the model. As this error increases to 100\%, it is unclear whether radiation pressure is capable of driving the wind. This is a particular concern for the less-luminous stars with high mass-loss rates: x03, V18 and V13. While expansion velocities of 3--4 km s$^{-1}$ are not unreasonable when compared to Halo carbon stars \citep{GOL97,LZM+10}, the lower velocities we derive for those stars losing less mass become increasingly unreliable due to this error. \citet{BW91} predict that stars with metallicities below [Fe/H] = --1 cannot drive winds via radiation pressure, and must rely on acoustic (pulsation) driving to eject mass from the star (see also \citealt{Bowen88}). At [Fe/H] = --0.7, 47 Tuc is near this hypothetical limit, meaning we must consider acceleration by pulsation and other sources of energy.

The velocity amplitude of the photosphere in these variables ranges from 4--23 km s$^{-1}$, with V1, V2 and V3 showing line doubling in near-IR and/or line emission in the optical due to the propagation of shocks in the atmosphere (Table \ref{MdotTable}; \citealt{LWH+05}; \citealt{MvL07}). While this velocity greatly exceeds our calculated wind velocity of $\sim$4 km s$^{-1}$, the bulk material does not necessarily attain this velocity. Nevertheless, the presence of shocks in the extended atmosphere implies that energy is being transferred to larger radii, and implies that this energy is available for driving mass loss.

A comparison between the presence of circumstellar dust and pulsation reveals some interesting results. All the stars listed in Table \ref{MdotTable} are variable: Lee\,1424 = ACVS\,002331-7222.6, while x03 shows a variation of $\delta K_{s} \approx 0.15$ mag (N.\ Matsunaga, private communication). Conversely, 22 of 43 variable stars show at least probable evidence of IR excess. Dust-producing stars therefore appear to be (in 47 Tuc) exclusively variable, and variable stars tend to be dust producing. In both Figures \ref{HRDVarsFig} \& \ref{PLFig} there appears no easy way of discerning RGB-pulsators from AGB-pulsators: the outliers in Figure \ref{HRDVarsFig} are such due to crowding in the cluster core (Paper III), while the pulsation periods in Figure \ref{PLFig} do not show the bimodality that one might expect from having RGB and AGB stars at different masses. However, a bimodality may exist in mass-loss rate with a transition occurring around 2500--3000 L$_\odot$ (Figure \ref{MdotRelFig}; cf.\ a similar possible bimodality in the gas-production rate in \citet{MvL07}). The five stars with elevated mass-loss rates are also those with the highest pulsation velocity amplitudes and among those with the longest pulsation periods. As the presence of pulsation and its amplitude both appear to correlate with mass-loss rate, an obvious conclusion would be that pulsation energy is dissipated in the extended atmosphere and helps drive the wind, though we caution that the strongest variables are also the most-luminous stars with the highest capacity for driving a wind through radiation pressure.

Further evidence that pulsation sets the mass-loss rate can be drawn from the apparent invariance of mass-loss over the narrow range of temperatures we probe (Figure \ref{MdotRelFig}). The photospheric temperature sets the radius at which dust can condense and therefore (if other parameters are invariant) the density of the wind. Higher-density winds lead to more-efficient dust formation, as the timescale for collision between dust-forming molecules is shorter. An invariance of mass-loss rate with temperature thus means that the formation of dust has little impact on the mass-loss rates present, providing further evidence that the mass-loss rate is `hard-coded' into the wind before the dust-formation radius. If this is true, the r\^ole of radiation pressure on dust may simply be to affect the outflow velocity of the wind.

If the wind is acoustically driven by pulsations, then the wind velocity and thus mass-loss rate is liable to be several times greater than the values listed in Table \ref{MdotTable}, since these sources of energy are not included in our model. This poses potential problems, as mass-loss rates this high would lead to the dissipation of the star's atmosphere on much shorter timescales than is inferred from star counts. This problem may be solved if one invokes different parameters for dust in the wind, which could lead to mass-loss rates much smaller than the listed values (see \S\ref{DustCompSect}).

\subsection{Evolutionary problems with our mass-loss rates}
\label{DustEvolSect}

In Paper III, we discussed the integrated RGB mass loss for a typical cluster star. We noted it was likely $\lesssim$0.22 M$_\odot$ and possibly, to first order, invariant with metallicity. This invariance would further suggest that the primary driver of RGB mass loss is not related to dust production. We had also estimated in Paper III, based on the cluster's luminosity function, that one star reaches the AGB tip every 80 kyr. 

These figures, however, do not correlate with our derived mass-loss rates. We find here that the cluster's $\sim$22 dust-producing stars are losing a total of $\approx$1.2 $\times 10^{-5}$ M$_\odot$ yr$^{-1}$ of dust-containing wind. Above the RGB tip ($\approx$2000 L$_\odot$), where dust production is ubiquitous, 13 stars are producing $1 \times 10^{-5}$ M$_\odot$ yr$^{-1}$. This implies that a staggering $\approx$0.8 M$_\odot$ is lost in the $\approx$1 Myr which a star spends on the AGB above the RGB tip. Clearly, this is impossible, as this is the entire mass of the star. This argues for a mass-loss rate of order 4--8 times lower than we measure. If additional wind driving mechanisms are important, this factor will increase further.

Mass loss via dust-containing winds between 1000 and 2000 L$_\odot$ can be similarly determined. Here, nine out of the 56 stars in this luminosity range are producing $\sim$2 $\times 10^{-6}$ M$_\odot$ yr$^{-1}$, implying 0.16 M$_\odot$ is lost over a 4.5 Myr timescale. If one assumes that mass loss from these stars is also lower than measured, and that some or all of these stars are AGB stars, this implies that mass loss via non-dusty winds may be of similar or even greater importance in removing the estimated $\lesssim$0.22 M$_\odot$ from the star between the middle of the RGB and middle of the AGB.

At the rates implied in Table \ref{MdotTable}, the iron in the atmosphere of V1 would be entirely ejected within 10\,000--40\,000 years, depending on the remaining envelope mass (0.08--0.3 M$_\odot$). V8, V2 and V3 would similarly disperse the iron in their atmospheres within timescales of $<$90\,000 years. The Padova and Dartmouth isochrones used in Paper III predict that these stars (especially V2 and V3) still have several hundred kyr left on the AGB, implying that our iron mass-loss rates are (again) a factor of several too high.

\subsection{Constraining the wind parameters}
\label{WindParamSect}

\begin{center}
\begin{table}
\caption{Maximum elemental abundances in dust}
\label{PsiTable}
\begin{tabular}{l@{}r@{\ }r@{\ }r@{\ }r@{\ }r@{\ }r}
    \hline \hline
Element or	& \multicolumn{5}{c}{Relative number density}					\\
compound	& \multicolumn{1}{c}{At}	&\multicolumn{1}{c}{CO}	&\multicolumn{1}{c}{MgSiO$_3$}	&\multicolumn{1}{c}{Fe \& Ni}	&\multicolumn{1}{c}{Final}	\\
\ 		& \multicolumn{1}{c}{star}	&\multicolumn{1}{c}{forms}&\multicolumn{1}{c}{\llap{c}ondense\rlap{s}}	&\multicolumn{1}{c}{\llap{c}ondens\rlap{e}}	&\multicolumn{1}{c}{values}	\\
   \hline
H		&1000000&1000000&1000000&1000000&1000000\\
He		& 81970 & 81970 & 81970 & 81970 & 81970 \\
O		& 155  	& 59   	& 26   	& 26   	& 21	\\
C		& 96   	& 0	& 0	& 0	& 0	\\
Mg		& 14.1	& 14.1	& 3.2	& 3.2	& 0	\\
Si		& 11.0	& 11.0	& 0	& 0	& 0	\\
Fe		& 6.3 	& 6.3 	& 6.3 	& 0	& 0	\\
Ni		& 0.6 	& 0.6 	& 0.6 	& 0	& 0	\\
Ti		& 0.03 	& 0.03 	& 0.03 	& 0.03	& 0	\\
Cr		& 0.15 	& 0.15 	& 0.15 	& 0.15	& 0	\\
Mn		& 0.10 	& 0.10 	& 0.10 	& 0.10	& 0	\\
Ca		& 0.4	& 0.4	& 0.4	& 0.4	& 0	\\
Al		& 1.0	& 1.0	& 1.0	& 1.0	& 0	\\
   \hline
CO		& 0	& 96	& 96 	& 96	& 96   \\
MgSiO$_3$	& 0	& 11.0	& 11.0	& 11.0	& 11.0 \\
Metallic iron	& 0	& 0	& 6.3	&  6.3	&  6.3 \\
Metallic nickel	& 0	& 0	& 0.6	&  0.6	&  0.6 \\
Other oxides	& 0	& 0	& 0	&    0	&  4.2 \\
   \hline
Element or 	& \multicolumn{5}{c}{Relative mass density}\\
compound	&	&	&	&	&	\\
   \hline
H		&1000000&1000000&1000000&1000000&1000000 \\
He		& 325500& 325500& 325500& 325500& 325500 \\
CO		& 0	& 2675	& 2675 	& 2675	& 2675 \\
Others		& 5846 	& 3192	& 2100	& 1716	& 1482 \\
\emph{Total gas}&\emph{1331400}&\emph{1331400}&\emph{1330300}&\emph{1329900}&\emph{1329700} \\
   \hline
MgSiO$_3$	& 0	& 0	& 1101	& 1101	& 1101 \\
Metallic iron	& 0	& 0	& 0	&  352	&  352 \\
Metallic nickel	& 0	& 0	& 0	&   35	&   35 \\
Other oxides	& 0	& 0	& 0	&    0	&  222 \\
\emph{Total dust}&\emph{0}	&\emph{0}&\emph{1101}&\emph{1488}&\emph{1710}\\
   \hline
$\psi_{\rm max}$& 	&	&	&	& 1/778 \\
   \hline
\end{tabular}
\end{table}
\end{center}

In \S\ref{WindAccSect}, we find radiation pressure may not be the dominant driver of winds from metal-poor stars, and is perhaps incapable of driving them in some cases. The simplest solution is to provide momentum from another source, such as stellar pulsations, in order to increase the wind velocity. Unfortunately, this also has the effect of increasing the mass-loss rate, which is already higher than stellar evolutionary theory can accommodate (\S\ref{DustEvolSect}). While we expect another momentum source to be present in the most-metal-poor stars (simply because radiation pressure on dust is that inefficient), the limits implied by the mass-loss rates imply this is not the case in 47 Tuc.

A solution to this dilemma should address both these problems by decreasing the mass-loss rate without significantly decreasing the wind velocity. Examining Table \ref{MdotErrorTable}, the simplest change would be to bring the dust-to-gas ratio closer to unity, which would decrease the mass-loss rate and increase the wind velocity. Alternatively, we may consider measures that would increase the wind velocity without appreciable increasing the mass-loss rate, such as changing the size distribution of grains.

The dust-to-gas ratio, $\psi$, given in Eq.\ (\ref{PsiEqn}), assumes that the dust fraction scales with metallicity (cf.\ \citealt{vanLoon00}). This will hold true if every dust species condenses with the same efficiency. Na\"ively, one may imagine that the fraction of each element condensed would decrease toward low metallicity simply because there is less opportunity for metal-metal reactions to take place in a metal-poor environment. From our conventional understanding, it is difficult to see how the dust-to-gas ratio in the wind could be substantially closer to parity.

We can, however, calculate a maximum amount of dust per unit gas volume, by taking the major dust-forming elements (O, C, Fe, Si, Mg, Ca, Al) plus some other potential dust-forming metals (Ni, Ti, Cr, Mn) and determining how much dust they can form. For 47 Tuc, we assume the abundances of most elements are scaled from the solar values of \citet{AGSS09} by [Z/H] = $-0.45$ dex, following \citet{ABBO+05}. We take the following exceptions, however: [Fe/H] = --0.7 dex \citep{Harris96}; [He/H] $\approx$ --0.02 dex \citep{DVL89}; [O/Fe] = +0.2 dex \citep{DOLG+10}; [Mg/Fe] $\approx$ [Si/Fe] $\approx$ [Ti/Fe] = +0.23 dex and [Ca/Fe] = 0.0 dex \citet{ABBO+05}. Table \ref{PsiTable} shows the stages of this calculation, by which we conclude that $\psi \lesssim 1/778$. Here we make the simplistic assumptions that all dust-forming elements start out in the gas phase, that all carbon forms CO, that all silicon forms enstatite (MgSiO$_3$), that iron and nickel condense entirely as pure metals, and that the leftover metals all form their respective oxides. Chemical reactions must be calculated on the basis of number density, but these must be converted to mass densities in order to calculate a dust-to-gas (mass) ratio. For reference, the same calculation using solar metallicities arrives at $\psi \lesssim 1/235$.

This approach neglects several points. Carbon may be enhanced in these stars due to dredge-up\footnote{While efficient dredge-up is normally considered to be limited to more-massive stars ($\gtrsim$ 2 M$_\odot$), carbon-enhanced stars of $\sim$0.8 M$_\odot$ exist in globular clusters (e.g.\ \protect\citealt{vLvLS+07}).}, but unless it is sufficiently abundant (C/O $\gtrsim$ 0.75) to bind all the remaining free oxygen, this does not affect the end result. We also neglect water condensation, which could use up free oxygen, but this would serve to decrease $\psi$. Forsterite (Mg$_2$SiO$_4$) may condense out in preference to enstatite, and calcium and aluminium may form into silicates before enstatite, but these would also decrease $\psi$ where they affect it at all. The formation of FeO and iron-rich silicates would increase $\psi$, but it seems that only a small percentage of iron condenses into non-metallic forms in these stars. Our stated value of $\psi_{\rm max}$ is therefore a relatively robust maximum value.

\subsection{Dust condensation efficiency}
\label{DustCondSect}

Table \ref{MdotTable} suggests that the number density of iron grains to other dust species in the winds is very high (typically 10:1, where it can be measured), but the analysis shown in Table \ref{PsiTable} would suggest it is of order 1:2 if all dust is condensed. This would suggest that the fraction of silicate dust that is condensed is quite low ($\sim$5\%), however we have not yet accounted for two possibilities: that the temperatures of the grain populations may be limited, or that the size distributions (specifically the surface-area-to-volume ratios) of silicates and iron may be different.

Differing grain temperatures between dust species may exist when the wind is sparse enough that dust-gas collisions cannot equalise the thermal energy. This would mean that more-opaque grains, like iron, are warmer; more-transparent grains, like silicates are colder for a given radius, and could explain the difficulties we have in fitting both the metallic-iron continuum, and the 10- and 20-$\mu$m silicate bands simultaneously (\S\ref{MdotSect}). More significantly, however, if thermal energy cannot be equalised via collisions, this will lead to a significant velocity drift between the gas and dust components of the wind. This will substantially decrease the projected mass-loss rate, while the dust outflow velocity will also increase accordingly. If the gas outflow velocity is below the turbulent velocity in the wind, this may allow gas to preferentially fall back, making the star more metal poor. Post-AGB stars and planetary nebulae are known to be metal-poor for similar reasons (e.g.\ \citealt{Cami02,MvWLE05,DIRMV09,JP10}), but it is not clear that this process is happening while the stars are still on the AGB where the outflow velocities are lower and the wind densities at the dust-formation radius are greater.

Conversely, if dust grain temperatures are equalised and the formation temperatures of silicates and iron are roughly similar, we may examine the effect of different grain sizes by taking two limiting cases. If we take the limiting case that both silicates and iron grains follow the same continuous distribution of ellipsoids, the silicate condensation fraction is $\sim$5\%, and we arrive at a value of $\psi \approx 1/2000$ to 1/3000 if all the iron is condensed. This would increase our derived $\dot{M}$ values by $\approx$60\% while decreasing $v_{\infty}$ by $\approx$30\%. This is the opposite effect to that needed to reconcile our observations with evolutionary theory, therefore we consider this scenario unlikely. In the converse limit, silicate and iron dust has condensed efficiently, but the iron grains have average radii 20$\times$ smaller. This increases $\dot{M}$ by a moderate (30--60\%) amount, but increases $v_{\infty}$ by much more (100--200\%), depending on the exact size distribution. If one is prepared to accept that iron grains grow preferentially in one direction, e.g.\ if magnetic effects are important, needle-like cylinders would be created (cf.\ \citealt{KdKW+02}). This would lead to a larger $v_\infty$, but in this case without an increased $\dot{M}$. However, if identical fractions of silicates and iron are condensed, the length/radius ratio of the iron grains would need to be a staggering $\approx$1200, meaning that the grains must be smaller as well. For example, a population of grains 5$\times$ smaller with a length/radius ratio of 10 allows a similar fraction of silicates and iron to be condensed with only a slight increase in $\dot{M}$.


There is no obvious mechanism that would cause iron to be \emph{preferentially} condensed at low metallicities, though magnetic effects and/or the basic problem of forming grains may be important (an iron grain only needs two iron atoms to bond, whereas a silicate grain requires a series of chemical reactions to form MgSiO$_3$). Higher partial pressures of iron may also assist here, though it is not clear why this would happen in a metal-poor environment.

Note that in all cases above, we have assumed grains are homogeneous: our calculation of the wind parameters in \S\ref{MdotSect} assumes all grains are a uniform mix of the included species, while our discussion above takes the opposite limit: where grains of each species exist separately in pure form. In reality, grains aggregate and form inclusions inside each other, which complicates the above arguments. In general, however, the evolutionary constraints provided in \S\ref{DustEvolSect} would suggest that dust condensation efficiency is high compared to solar-metallicity stars (suggesting $\psi > 1/1000$), and the iron grains are smaller and/or more-elongated than both the silicate grains and the assumed MRN distribution. A combination of these effects is likely required to reduce $\dot{M}$ by the factor of 4--8 required without reducing the outflow velocities, $v_\infty$, we derive (Table \ref{MdotTable}).

\subsection{Composition and radial driving of dust}
\label{DustCompSect}

\begin{figure}
\includegraphics[width=0.35\textwidth,angle=-90]{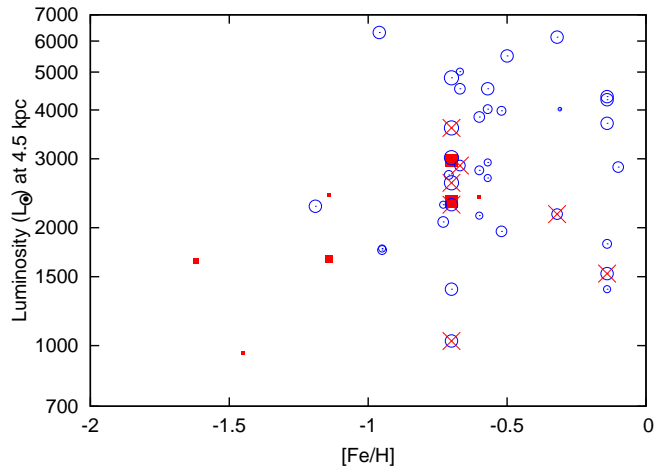}
\caption{Dust type and mass-loss rate as a function of luminosity, showing additional data sourced from \protect\citet{MSZ+10} and Paper I. Filled, red squares have only metallic iron dust, hollow blue circles also show silicates, while crosses show the stars with obvious 13.1 and 20-$\mu$m features. Dot size corresponds to \emph{estimated} mass-loss rate.}
\label{FeHFig}
\end{figure}

Where we are able to determine it, the composition of circumstellar dust and its variation with luminosity is similar to that we see in other clusters in \citet{MSZ+10} (see also Figure \ref{FeHFig}). The evolution of dust composition in 47 Tuc broadly agrees with that seen in other clusters at this metallicity. Metallic iron appears to be ubiquitous around dust-producing stars, while amorphous silicates are produced by most stars without any obvious dependance on luminosity.

The published spectra of several stars (V4, V8, V13, and weakly in V21; see \citealt{LPH+06}) show sharp features around 9.7, 11.3, 13.1, 20 and 32 $\mu$m. \citet{MSZ+10} attribute the 20-$\mu$m feature to iron-rich w\"ustite (Mg$_{1-x}$Fe$_{x}$O), where $x \approx 1$). Recent work by Niyogi, Speck \& Onaka (submitted) suggests that all these features, with the exception of the 13.1 $\mu$m feature (which is probably due to an aluminium oxide --- \citealt{SKGP03}), can be explained by iron-rich crystalline silicates. Here, the 20-$\mu$m feature is still produced by an Fe-O bond, but it exists in a large crystalline lattice of fayalite (Fe$_2$SiO$_4$) or ferrosilite (FeSiO$_3$).

Iron-rich silicates are not expected to form in circumstellar environments, as their condensation temperatures are below that of both magnesium-rich silicates and metallic iron \citep{GS99}. If iron-bearing crystalline silicates are a significant component of some winds, they must exist alongside metallic iron and amorphous silicate species. One way this could occur is if cooling of the wind proceeds at a faster rate than dust condensation, which may occur in the tenuous winds of low-luminosity, metal-poor stars. Figure \ref{FeHFig} shows that globular cluster stars which show a 20-$\mu$m feature may be confined to luminosities below $\sim$4000 L$_\odot$. It is difficult to say much about their metallicity dependance due to insufficient data, especially given silicate production appears to decline below [Fe/H] $\sim$ --1.

Interestingly, the winds found in such stars (x03, V13 and V18) appear to be quite cool. In more-luminous stars, we find wind temperatures to be higher ($\sim$1000 K). The condensation of metallic iron before silicates remains a mystery: for pressures where iron condenses around $\sim$1000 K, magnesium-rich silicates should have already formed closer to the star (e.g.\ \citealt{GS99}, their Figure 2). Significant departures from LTE, or unusually-high partial pressures of iron (caused, perhaps, by magnetic confinement or clumping in the wind) may offer ways to circumvent this problem, but to invoke them as solutions would be speculative without further data.

By whichever process metallic iron is created, if radiation pressure is effective in wind driving, metallic iron condensation should have a dramatic effect on the wind. Its condensation should cause a rapid acceleration of the wind, dropping the density and preventing other dust species from condensing. This may be what we see in moderately-luminous stars, such as V3, LW10 and V11. 

If radiation pressure is not an effective driver, the general absence of other dust species may be explained if silicates form using iron grains as seed nuclei. In this case, denser or more-metal-rich winds effectively condense silicates onto iron grains via collisions, while less-dense or metal-poor winds expand before much condensation occurs, leaving bare iron grains. This scenario may explain the iron grains found in the interplanetary dust particles known as GEMs (glasses with embedded metal and sulphides; \citealt{Bradley94,Martin95}).

In the most-evolved stars (V1, V8, V2 and V4), however, we do see silicate emission making an increasingly-strong presence. Meanwhile, there is a hint that the temperature in the dust formation zone (as traced by metallic iron condensation) may be decreasing. As noted in \S\ref{DustCondSect}, different dust species may also have different temperatures. More detailed modelling of the winds of these stars, allowing for thermal decoupling of the grain species, is required if we are to speculate on factors which cause this temperature change. It could, however, be linked to stronger pulsation of the star and opacity shielding within the wind. In a cooler wind, however, silicates once again condense out in increasing preference to metallic iron, which contradicts our tentative finding that the silicate grains are cooler than the iron grains in these stars. 


\section{Conclusions}
\label{SectConc}

In this work, we have used spectral energy distributions and infrared spectroscopy to establish mass-loss rates and investigate mineralogy around the dust-producing stars in 47 Tuc. We summarise our conclusions as follows:
\begin{list}{\labelitemi}{\leftmargin=1em \itemsep=0pt}
\item Mass loss in the cluster is dominated by its most luminous stars (V1, V8, V2, V3 and V4), however substantial mass loss appears to be taking place in moderate-luminosity stars too (x03, V18, V13).
\item The total mass-loss rate is estimated to be $\sim$1.2 $\times$ 10$^{-5}$ M$_\odot$ yr$^{-1}$, mainly in the form of metallic iron, but this is probably over-estimated by a factor of 4--8 due to unconstrained properties of the dust grains.
\item It remains unclear whether radiation pressure on dust is capable of driving a wind alone, due to the low projected terminal velocities of the stellar winds. The existence of circumstellar dust appears to correlate with the presence of stellar pulsation. We theorise that such pulsation may inject energy into the wind acoustically.
\item A dust-to-gas ratio closer to unity and smaller or more needle-like iron grains are suggested as methods to decrease the total mass-loss rate while allowing for an increase in the terminal velocity.
\item Variations in the wind chemistry broadly concur with those found in \citet{MSZ+10}, with silicates found preferentially in high-luminosity, high-metallicity winds, while iron is found preferentially in low-luminosity, low-metallicity winds. Iron may condense preferentially at low metallicity. Alternatively, iron grains may be coated by silicates in dense, metal-rich winds, but this coating process may not be effective in winds around lower-luminosity and metallicity stars.
\end{list}

\vspace{5 mm}
\noindent
{\bf Acknowledgments:} We are grateful to Thomas Lebzelter for sharing his original data with us. This paper uses observations made using the \emph{Spitzer Space Telescope} (operated by JPL, California Institute of Technology under NASA contract 1407 and supported by NASA through JPL (contract number 1257184)); observations using \emph{AKARI}, a JAXA project with the participation of ESA; and data products from the Two Microns All Sky Survey, which is a joint project of the University of Massachusetts and IPAC/CIT, funded by NASA and the NSF.


\end{document}